# Dynamics of the brain extracellular matrix governed by interactions with neural cells

Ivan A. Lazarevich[1,2], Sergey V. Stasenko[1], Maia A. Rozhnova[1],
Evgeniya V. Pankratova[1],
Alexander E. Dityatev[3], and Victor B. Kazantsev[1]
[1]*Lobachevsky State University of Nizhni Novgorod,*
*23 Gagarin ave., 603950 Nizhny Novgorod, Russia*
[2]*École Normale Supérieure, Paris Sciences et Lettres University,*
*Laboratoire de Neurosciences Cognitives, Group for Neural Theory, Paris, France and*
[3] *German Center for Neurodegenerative Diseases,*
*Group for Molecular Neuroplasticity, Magdeburg, Germany*

Neuronal and glial cells release diverse proteoglycans and glycoproteins, which aggregate in the extracellular space and form the extracellular matrix (ECM) that may in turn regulate major cellular functions. Brain cells also release extracellular proteases that may degrade the ECM, and both synthesis and degradation of ECM are activity-dependent. In this study we introduce a mathematical model describing population dynamics of neurons interacting with ECM molecules over extended timescales. It is demonstrated that depending on the prevalent biophysical mechanism of ECM-neuronal interactions, different dynamical regimes of ECM activity can be observed, including bistable states with stable stationary levels of ECM molecule concentration, spontaneous ECM oscillations, and coexistence of ECM oscillations and a stationary state, allowing dynamical switches between activity regimes.

## I. INTRODUCTION

Understanding principles and mechanisms of information processing in the central nervous system is among the main objectives of neuroscience. For a long time, the main role in this process was assigned to neurons. Recent experiments have shown that, in addition to neurons, an important role in the processing of information also belongs to glial cells and the ECM [1–4].

In a number of experimental studies, it was shown that the ECM molecules are capable of modulating the efficiency of synaptic transmission and neuronal excitability. It is assumed that these mechanisms play a key role in the homeostatic regulation of neuronal activity at relatively long time scales [1, 2]. The homeostatic form of plasticity caused by ECM allows the preservation of neuronal cells, preventing pathological hypo- and hyperexcitation of neurons, which can lead to neuronal dysfunction and cell death. For example, such a known effect as the synaptic scaling observed in the experiment allows neurons to maintain neuronal firing rate in a certain range in response to various alterations of afferent inputs [5, 6]. The change in the concentration of ECM receptors on postsynapses (integrins) led to a change in the expression of AMPA receptors, which eventually changed the efficiency of synaptic transmission [1]. Another cascade of regulation involves changing the $Ca^{2+}$ influx into neurons through interaction between hyaluronic acid and L-type calcium channels (L-VDCC) [7]. Regulation of ECM concentration is implemented not only via the control of synthesis and secretion of ECM molecules into the extracellular space, but also by the activity of proteases (tissue plasminogen activator, plasmin, matrix metalloproteinases 2 and 9, aggrecanases 1 and 2, neuropsin and neurotrypsin), released pre- and postsynaptically and cleaving the ECM molecules. As seen in experimental studies on hippocampal interneurons, ECM-neuron interactions involving neuronal Kv channels effectively lead to modulation of the action potential generation threshold, so that a deficit in ECM promotes firing of interneurons [8–10]. On the other hand, recent experimental findings for pyramidal neurons suggest that less spikes are generated after ECM attenuation due to activation of SK channels [11]. Thus, the considered regulations, mediated by the activity of ECM molecules, can lead to excitation or inhibition of neuronal activity. In this study we aim to investigate, using a mathematical model of ECM-neuronal interactions, how different regulation mechanisms involved in neuron-matrix interactions shape the dynamics of ECM production and degradation.

A phenomenological model describing the homeostatic regulation of neuronal activity by ECM molecules was first proposed by Kazantsev and others [2]. The model employed kinetic activation-function description of ECM activity and predicted that modulation of synaptic transmission and spiking threshold may lead to the appearance of two stable levels of homeostatic neuronal activity.

In the present work, we consider how prevalence of particular mechanisms of ECM-neuronal interactions might determine the dynamics of ECM concentration levels. We demonstrate that bistability with stable stationary states may be observed regardless of the polarity of ECM influence on neurons – it may be either inhibitory or excitatory. However, in the case when ECM has inhibitory effect on neuronal activity, we predict that bistability is dependent on the activity of proteases, while it is not the case when ECM-neuronal influence is excitatory. Excitatory ECM-neuron feed-back signals may also lead to spontaneous oscillations of ECM molecule concentration, which can coexist with a stable stationary state.

## II. MATHEMATICAL MODEL OF ECM ACTIVITY

The processes of ECM synthesis and degradation in a neuronal network are described with a phenomenological approach developed in [2]. Description of neural activity is in accordance with the mean-field Wilson-Cowan type model [12]. Due to the fact that the characteristic timescales of neural dynamics are significantly shorter than those of ECM molecule concentration changes, we set the mean firing rate of the neural population equal to the stationary value, which is a function of the ECM molecule concentration $Q = Q_{\inf}(Z)$. We assume here that only a single stationary value of the mean firing rate, e.g. we do not consider bistability induced by E-I interactions in the Wilson-Cowan model [12]. Depending on the polarity of ECM-neuron interactions, the function $Q_{\inf}(Z)$ can be either monotonically increasing or decreasing. The key variables describing ECM activity are the ECM concentration $Z$, the concentration of ECM receptors $R$, and the concentration of proteases $P$. The dynamical model consists of the following equations

$$\frac{dZ}{dt} = -(\alpha_Z + \gamma_P P)Z + \beta_Z H_Z(Q_{\inf}(Z)). \quad (1)$$

$$\frac{dP}{dt} = -\alpha_P P + \beta_P H_P(Q_{\inf}(Z)). \quad (2)$$

$$\frac{dR}{dt} = -\alpha_R R + \beta_R H_R(Q_{\inf}(Z)) \quad (3)$$

Here the activation functions $H_{Z,P,R}$ all assumed to have a sigmoid form. An increase in the protease concentration $P$ is assumed to be linearly related to the speed of ECM degradation $\alpha_Z^* = \alpha_Z + \gamma_P P$. If ECM-neuronal interactions involve effects of synaptic scaling [13], then stationary neuronal firing rate might also depend on the concentration of postsynaptic ECM receptors. We assume that the resultant extent of the synaptic scaling effect is proportional to the product of ECM molecule concentration and ECM receptor concentration $ZR$, since production of ECM molecules and receptors is assumed to be a statistically uncorrelated process. In the case of synaptic scaling, it was shown [2] for a Hodgkin-Huxley-type model that the resultant stationary firing rate $Q_{\inf}$ can be approximated by a linear function of $ZR$. The timescales of ECM receptor dynamics are at least an order of magnitude shorter than those of ECM molecules and receptors in the original model [2], so that variable $R$ can be approximated by its steady-state value $R_{\inf}(Q)$. For other ECM-neuron

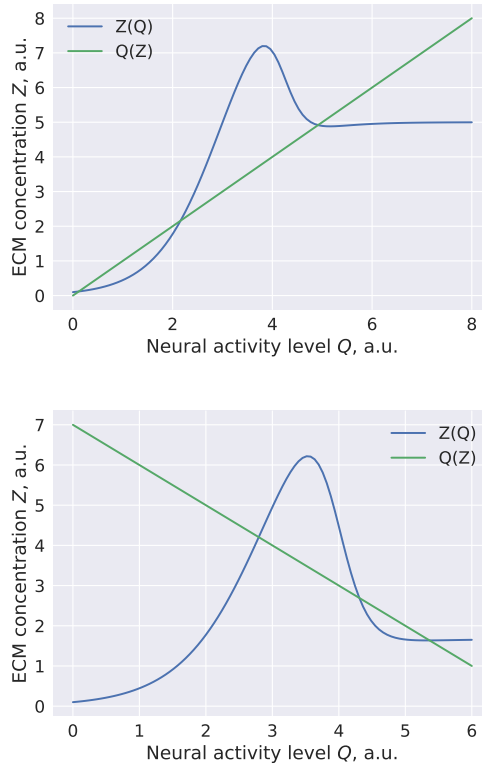

FIG. 1: Examples of equilibrium curves corresponding to equations (4)-(6) in the $(Z, Q)$ phase plane for the case of (top) excitatory ECM-neuron interaction and (bottom) inhibitory ECM-neuron interaction. Both fugures show existence of bistable solutions regardless of the polarity of ECM-neuron interactions

interaction mechanisms there is no dependence on the ECM receptor concentration $R$, as shown in further sections. In any case the dynamical system might be reduced to a two-dimensional one, so that it is rather analytically tractable.

## III. ECM BISTABILITY

Let us consider the case when the ECM-neuron interaction feedback loop involves modulation of either Kv channels (inhibitory ECM effect) or SK potassium channels (excitatory ECM effect). In these cases ECM-neuron interactions are independent of postsynaptic ECM receptor concentration $R$, since the regulation mechanism involves modulation of somatic membrane receptors of the neuron. Hence, the stationary firing rate of neurons depends only on the ECM concentration.

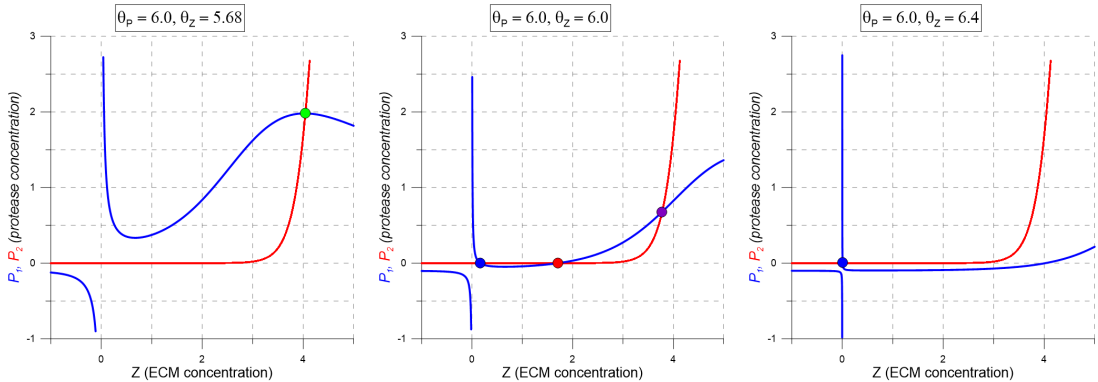

FIG. 2: $Z$, $P$ equation nullclines for different values of the ECM production threshold $\theta_Z$

We assume the effect of ECM concentration on neuronal firing rate might be approximated by a linear dependence $Q_{\text{inf}} = Q_0 + \alpha_Q Z$. This is a fair assumption when AP firing threshold is being modulated by ECM [2], and we use the same description when neuronal firing is modulated through SK2 channel activation. We arrive at the following system of equations describing ECM dynamics:

$$\frac{dZ}{dt} = -(\alpha_Z + \gamma_P P)Z + \beta_Z \hat{H}_Z(Z) = \hat{Z}(Z, P) \quad (4)$$

$$\hat{H}_Z(Z) = \left(Z_0 - \frac{Z_0 - Z_1}{1 + \exp(k_Z^{-1}(Q_0 + \alpha_Q Z - \theta_Z))}\right) \quad (5)$$

$$\frac{dP}{dt} = -\alpha_P P + \beta_P \hat{H}_P(Z) = \hat{P}(Z, P) \quad (6)$$

$$\hat{H}_P(Z) = \left(P_0 - \frac{P_0 - P_1}{1 + \exp(k_P^{-1}(Q_0 + \alpha_Q Z - \theta_P))}\right) \quad (7)$$

Let us first qualitatively show that ECM concentration might be bistable in this system regardless of the sign of $\alpha_Q$. The equilibrium curves in the ECM-concentration firing rate phase plane $(Z, Q)$ are shown in Fig. 2. It is apparent that there are cases of bistability, which correspond to the line $Q_{\text{inf}} = Q_0 + \alpha_Q Z$ intersecting the $Z_{\text{inf}}$ curve in three points, two stable and one unstable stationary solutions, correspondingly. Note that depending on the sign of the $\alpha_Q$ parameter, which controls whether ECM influence on neurons is inhibitory or excitatory, the bistability effect is induced by different mechanisms. When the ECM-neuron interaction is excitatory, and hence the slope of the $Q_{\text{inf}}(Z)$ line is positive, there can exist bistable solutions regardless of whether the curve $Z_{\text{inf}}(Q)$ has a "bump" at intermediate values of $Q$. A monotonically increasing sigmoid form of $Z_{\text{inf}}(Q)$ (which corresponds to the absence of protease effect on ECM, e.g. $\alpha_P = 0$) would be enough to yield a set of bistable solutions. On the other hand, if the ECM-neuron effect is inhibitory (negative $\alpha_Q$), bistable solutions only exist in the presence of the bump in the equilibrium curve $Z_{\text{inf}}(Q)$. This bump occurs because when neuronal firing rate $Q$ increases, the synthesis of ECM molecules is upregulated, but the concentration of proteases $P$ increases as well, though at slightly higher values of the firing rate. Increase in protease concentration $P$ leads to ECM degradation, hence the equilibrium value $Z_{\text{inf}}$ is smaller at higher firing rates compared to the intermediate range of $Q$ values. The height of this bump is determined by the strength of protease-induced ECM degradation (value of $\alpha_P$).

In biophysical terms, we predict that if the prevalent regulation cascade determining ECM-neuronal interactions restrains neuronal excitability, then ECM bistability can only be implemented if proteases demonstrate a strong effect on ECM degradation. If ECM-neuronal interactions mainly neuronal excitability, the bistability effect does not depend on the strength of protease-ECM interaction and might be implemented even in the absence of protease-dependent ECM degradation.

## IV. HOMEOSTATIC ECM OSCILLATIONS

Let us more closely consider the case of excitatory ECM-neuron interactions ($\alpha_Q > 0$), for instance implemented through activation of neuronal SK2 channels, as seen experimentally. First, let us study the number and stability of equilibrium states of the Eq. system (5)-(7). We set all parameters fixed with the following values: $Q_0 = 5, \alpha_Q = 0.23, \alpha_Z = 0.0001 \text{ ms}^{-1}, \gamma_P = 0.001 \text{ ms}^{-1} \beta_Z = 0.01 \text{ ms}^{-1}, \alpha_P = 0.001 \text{ ms}^{-1}, \beta_P = 0.001 \text{ ms}^{-1}, \theta_P = 6, k_Z = 0.15, k_P = 0.05$. The free parameter we consider is the effective firing rate threshold for ECM production, $\theta_Z \in [5.7, 6.5]$.

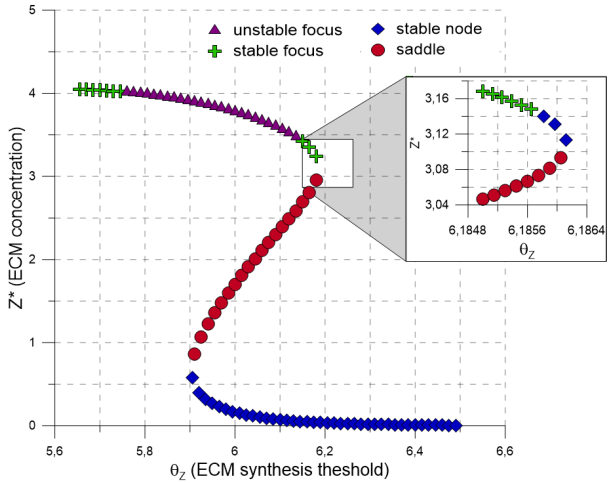

FIG. 3: Bifurcation diagram showing the change in the number and type of equilibrium points of the system (4)-(7) when the ECM production threshold value $\theta_Z$ is varied

The number of equilibrium points is determined by the number of intersections of the nullclines $\hat{Z}(Z,P) = 0$ and $\hat{P}(Z,P) = 0$. Fig. 2 shows the nullcline intersections for three different values of $\theta_Z$: $\theta_Z = 5.68$, $\theta_Z = 6$ and $\theta_Z = 6.4$. It is apparent that changes in $\theta_Z$ only influence the curve $\hat{Z}(Z,P) = 0$ while the $\hat{P}(Z,P) = 0$ curve stays the same. With increasing $\theta_Z$ the upper part of the $\hat{Z}(Z,P) = 0$ curve goes down relative to the $\hat{P}(Z,P) = 0$ curve. For $\theta_Z = 6$ three intersection points exist, which means that there exist two values $\theta_Z = \theta_Z^1 < 6$ and $\theta_Z = \theta_Z^2 > 6$ for which two curves touch each other, and points denoted blue and red coincide at $\theta_Z = \theta_Z^1$, while points denoted red and purple coincide at $\theta_Z = \theta_Z^2$. The system has three equilibrium points in the interval $\theta_Z \in (\theta_Z^1, \theta_Z^2)$.

To investigate the stability of these equilibrium points we calculate the Jacobi matrix

$$A = \begin{bmatrix} -(\alpha_Z + \gamma_P P) + A_{11} & -\gamma_P Z \\ A_{21} & -\alpha_P \end{bmatrix}$$

$$A_{11} = -\frac{\alpha_0 \beta_Z (z_0 - z_1) \exp(-k_Z^{-1}(Q_0 + \alpha_0 Z - \theta_Z))}{k_Z \left(1 + \exp(-k_Z^{-1}(Q_0 + \alpha_0 Z - \theta_Z))\right)^2}$$

$$A_{21} = -\frac{\alpha_0 \beta_P (z_0 - z_1) \exp(-k_P^{-1}(Q_0 + \alpha_0 Z - \theta_P))}{k_P \left(1 + \exp(-k_P^{-1}(Q_0 + \alpha_0 Z - \theta_P))\right)^2}$$

The roots of its characteristic equation determine the stability of equilibrium points Figure 3 shows the changes in the stationary $Z$ value when the ECM production threshold $\theta_Z$ is varied in the interval $[5.7, 6.5]$.

Different symbols in the figure denote different types of equilibrium points. In particular, the stable equilibrium regimes on the Z-shaped curve are denoted by green (stable focus) and blue (stable node) symbols. In the considered range of $\theta_Z$ the system (1) can demonstrate globally stable (monostable) equilibrium regimes, in partiular, for $\theta_Z < 5.68$ the only attractive manifold in the phase space is a stable focus, while for $\theta_Z > 6.19$ it is a stable node. The bistable behavior of the system caused by the coexistence of two stationary states is observed for $\theta_Z \in (6.15, 6.19)$.

In addition to stable stationary states, there might exist oscillatory regimes in the system as well, with corresponding limit cycles in the phase space of the system. Blue curves in Figure 4 demonstrate the minimal and maximal values, which $Z$ can achieve on the stable limit cycle for different values of $\theta_Z$. The red curves denote the same, but for the unstable limit cycle. In particular, for $\theta_Z \approx 5.685$ a stable and an unstable limit cycles appear as a result of a double-limit-cycle bifurcation. The phase portrait for $\theta_Z \approx 5.69$ is shown in Fig. 6b. In this figure and others the stable limit cycle is drawn with blue color, and the unstable limit cycle with red color. For $\theta_Z \approx 5.755$, as a result of an Andronov-Hopf bifurcation, the limit cycle turns into the equilibrium point, and the stable focus becomes unstable, as shown in Fig. 6c. In the $\theta_Z \in (5.755, 5.904)$ interval there is a single attractive manifold in the phase space, which is a stable limit cycle. It disappears through a saddle-node bifurcation at $\theta_Z \approx 5.904$. The two equilibrium states that appear as a result of the bifurcation (a stable node and a saddle) "walk away" with increasing $\theta_Z$ and, in particular, for $\theta_Z \approx 5.92$ the phase portrait of the system has a form as shown in Fig. 6d.

With a further increase in the value of $\theta_Z$, another limit cycle appears. Mechanism of its appearance is shown in Fig. 6e: for $\theta_Z = 6.1$ an unstable separatrix bends the stable separatrix outside; the separatrices get closer with increasing $\theta_Z$, and for $\theta_Z = 6.12$ the stable separatrix covers the unstable one. The change in relative situation of separatrices is taking place with a negative saddle value $\sigma = \lambda_1 + \lambda_2 < 0$. Therefore, a stable limit cycle has to appear, which is exactly what is observed: Fig. 6f shows the cycle which appears is born as a result of separatrix-loop saddle-node bifurcation with blue color. It is noteworthy that the amplitude of this oscillatory state is rather small and its generation depends on the initial conditions, since it is coexistent with a stable node in the phase space. This limit cycle is observed for $\theta_Z \in (6.11, 6.14)$. For $\theta_Z \approx 6.14$ the stable cycle turns into the equilibrium point and vanishes as a result of another Andronov-Hopf bifurcation. With a further increase of $\theta_Z$ the focus which changes stability as a result of the Andronov-Hopf bifurcation turns into a node (Fig. 6g) and disappears (Fig. 6h) as a result of another saddle-node bifurcation at $\theta_Z \approx 6.19$.

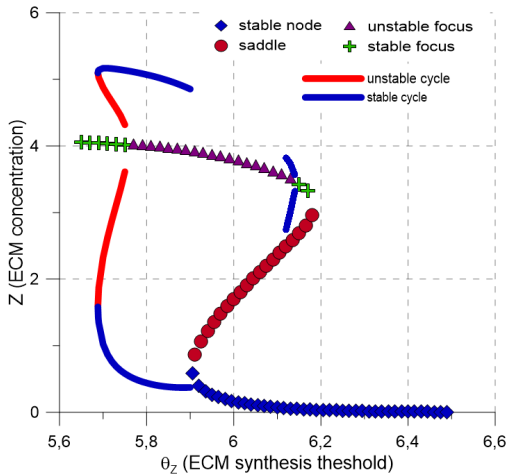

FIG. 4: Bifurcation diagram of the system (4)-(7) when $\theta_Z$ value is varied. Phase portraits of the system corresponding to various fixed values of $\theta_Z$ are shown in Figure 5.

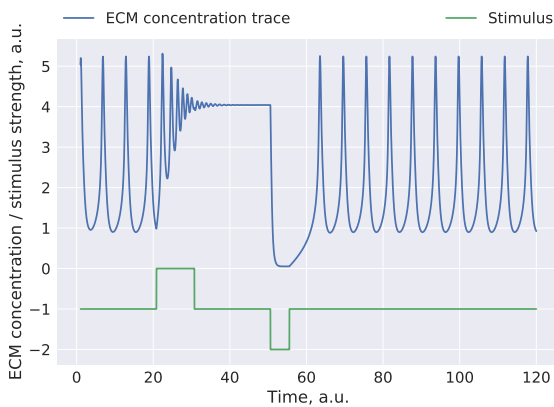

FIG. 5: Simulated ECM concentration trace under the conditions when the ECM-protease system exhibits coexistence of a stable limit cycle and a stable stationary state. Application of an external stimulus (e.g. a spontaneous increase or decrease in neural activity) may induce dynamical switches between activity states.

In summary, with the increasing value of ECM production threshold $\theta_Z$ we can observe two areas in the parameter space of the model, where oscillatory dynamics of ECM concentration levels might occur, either spontaneously (limit cycle is the only stable manifold in the phase space) or a result of external stimulation (if ECM concentration was initially in a stationary state).

If we look at bifurcation diagrams of the system as $\theta_Z$ changes at different fixed levels of $\theta_P$ - the protease production threshold, we can notice slight changes in the system's dynamics. Figure[add] shows possibility of coexistence of a stable limit cycle, a stable focus and a stable node for a certain value of $\theta_P$, whereas for a different $\theta_P$ value one can observe that the limit cycle born through an Andronov-Hopf bifurcation vanishes through another Andronov-Hopf bifurcation, instead of a separatrix-loop bifurcation. The nature of these ECM oscillations may be understood qualitatively - an increase in neuronal activity drives an increase in ECM concentration, and further release and activation of proteases that degrade ECM molecules, which in turn lowers neuronal activity. Proteases are less active at low neuronal activity levels, and the positive ECM-neuronal firing feedback loop drives the activity levels up again.

Figure 5 shows neuronal firing-induced switches between oscillations and a stationary ECM state. Spontaneous changes in the level of neural firing act as an effective stimulus to the ECM-proteases system, which might drive the system away from the locally stable manifold.

The physical timescale values of the observed ECM oscillations are quite flexible in our model, since the key assumption is that ECM dynamics is at least significantly slower as compared to neuronal dynamics. Experimentally observed changes in ECM concentration may be on the timescale of hours to days [1], and the exact relaxation time values in the model shall be further calibrated based on available experimental data.

## V. INFLUENCE OF ECM RECEPTOR DYNAMICS

In the case when the prevalent mechanism of ECM-neuron interactions is through synaptic scaling, the dynamics of ECM receptors might influence ECM dynamics in general. As mentioned above, typically the characteristic timescales of ECM receptor dynamics is significantly shorter than that of ECM molecules and proteases, so that $R$ can be replaced with the stationary value $R_{\inf}(Q)$. The dynamics of ECM receptors is, however, significantly slower than that of neuronal activity, so we can set $R = R_{\inf}(Q_{\inf})$. Assuming that the stationary firing rate level scales linearly with the product $ZR$, we arrive at

$$Q_{\inf}(Z) = \frac{Q_0 + \alpha_Q R_0 Z}{1 + \alpha_Q \alpha_R Z} \quad (8)$$

where we also introduced a linear approximation $R \approx R_0 - \alpha_R Q$. It is clear that the slope of ECM-neuronal interaction curve is now activity-dependent, decreasing with higher levels of neural activity. This might result in an activity-driven formation of bistable or oscillatory states of the ECM concentration.

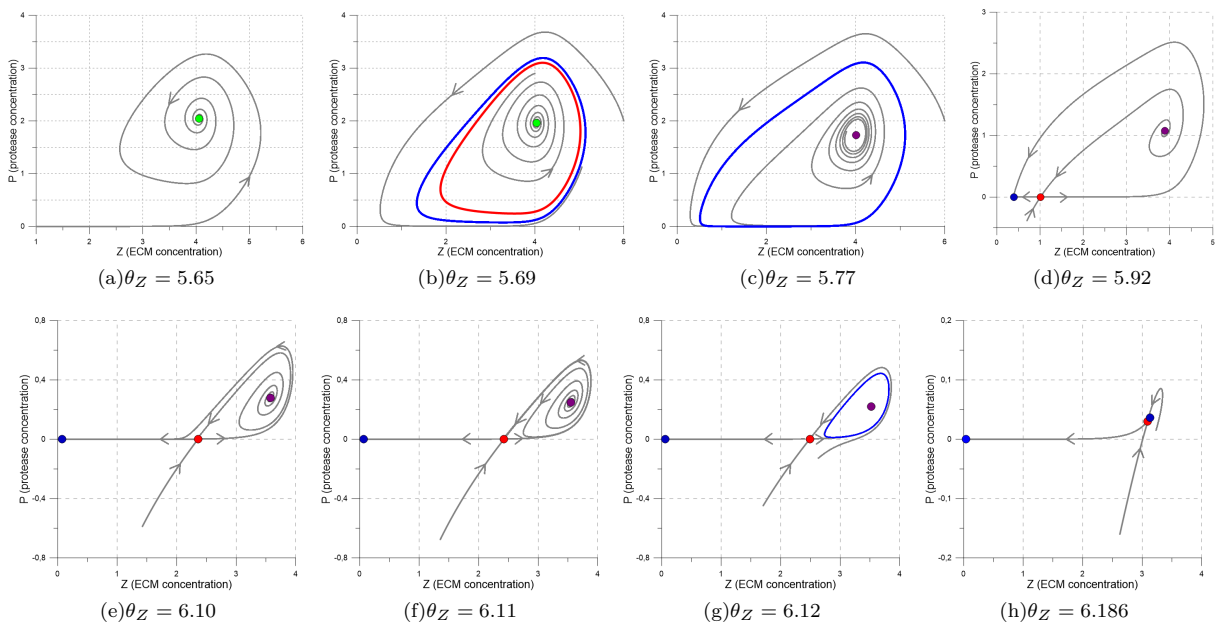

FIG. 6: Phase portraits of the ECM-proteases system at different values of $\theta_Z$. (a) Globally stable stationary state, (b) Coexistense of a stationary and an oscillatory state, (c) Globally stable oscillatory state, (d) Two stable stationary states, (e-g) Birth of the small-amplitude limit cycle as a result of the saddle-node bifurcation, (h) Two stationary stable states

Another limit case is when dynamics of ECM receptors slow even in comparison to characteristic timescales of ECM acitivity (e.g. the period of ECM oscillations), when the value of $R \approx R^*$ is approximately constant on the timescale of interest. In this case the analysis would be the same as in the case of SK2-channel mediated ECM-neuron interactions, with negligible activity-dependent changes to the system's dynamics.

## VI. CONCLUSIONS

In summary, we have investigated ECM molecule concentration dynamics in a mathematical model of ECM-regulated modulation of neural activity. The model is based on the following key assumptions: (a) synthesis of ECM molecules and ECM-degrading enzymes is controlled by the level of neuronal activity, (b) changes in ECM levels may in turn modulate neuronal activity, in either excitatory or inhibitory manner, depending on the prevailing mechanism of ECM-neuronal interaction. Mathematically, the model can be reduced to a set of two or three coupled differential equations, depending on the assumptions concerning the nature of ECM-neuronal interactions and characteristic timescales of postsynaptic ECM receptor production. Inhibitory effect of increased ECM levels on neural activity was observed to induce protease-dependent bistable dynamics, while the excitatory effect of ECM-neuronal interaction resulted in a richer repertoire of observable dynamical states. We found that, for the excitatory ECM-neuron interactions implemented through e.g. modulation of somatic SK-channels or through synaptic scaling, the ECM concentration levels may exhibit different activity regimes, ranging from neural firing-induced protease-independent switching between stationary states of the ECM concentration to spontaneous ECM oscillations, which might coexist with a stationary concentration level. In terms of neuronal activity, this means that there are different dynamical modes of ultra-slow firing threshold modulation or modulation of the power of the synaptic scaling effect. Development of more detailed network-wide models of neural activity subject to these ultra-slow modulations might reveal the functional effects by which changes in the extracellular matrix might shape the activity of neuronal circuits.

### Acknowledgments

This work was supported by the Ministry of education and science of Russia under project 14.Y26.31.0022